# A STUDY ON THE CONTROLLABILITY OF LITHIUM-ION BATTERIES

Preston Abadie[1], Donald Docimo[1]

[1]Texas Tech University, Lubbock, TX

**ABSTRACT**

*This work explores controllability and the control effort required for lithium-ion batteries. Battery packs have become a critical technology in both personal and professional applications as a means to store large amounts of energy. Management of cells in a pack becomes increasingly difficult though, with charging and discharging operations requiring more complex strategies due to parameter variations between the cells. There are numerous studies which develop effective estimation and control schemes to reduce the impact of the imbalances present in battery packs, but the receptiveness of the individual cells to these schemes is much less explored. This paper performs a nonlinear controllability analysis for experimentally parameterized cells. A connection is shown between the condition number of a battery's controllability matrix and the amount of control effort that battery will require. This reveals that if a cell's dynamics are poorly mathematically conditioned, it will require more time or higher power to control than one that is not. The controllability condition number of each cell's model is then determined both with new and aged parameters, and a sensitivity analysis shows that the cells' conditioning is equally impacted by all parameters. This offers insight into the increased control effort required for a battery as it ages and the culprit of said increase. The results of this analysis are then used to determine the best conditioned assemblies for a batch of cells with a mix of new and second-life parameters.*

## 1. INTRODUCTION

This paper analyzes how the mathematical conditioning of a battery cell's controllability index is impacted by changes in the cell parameters. Current energy production methods are racing to keep pace with the growing demand, but in addition to needing more generation, there is also a need to prevent wasted energy during low-demand hours. Energy storage has played a prevalent role in aiding renewable energy by storing excess power generated during peak production hours so that it is available for use at a later time [1], [2]. This has been increasingly done with lithium-ion battery packs. To ensure these packs are efficient and long-lasting, they are equipped with battery management systems (BMS) to manage the current received by cells or groups of cells [3], [4], [5]. This is implemented according to various control strategies intending to remove state and output heterogeneity as well as maintain safety limits [6], [7], [8]. These algorithms can prove effective; however, they are implemented without concern for how difficult it might be to control the cells' behavior. Batteries that are less controllable require more control effort, either through higher operating power or longer management scenarios.

Battery packs have numerous states which must be controlled in order to achieve performance metrics and lifetime requirements for the pack. The most popular type of control input for lithium-ion batteries is through cell current. This is necessary for charging and discharging the pack, with algorithms focusing on meeting power demands [9], [10] and keeping relevant states and outputs in safe operating ranges [8], [11], [12]. Another application of current control in lithium-ion battery packs is state balancing. This is done using hardware such as DC/DC converters in parallel with the cells and a common voltage bus to shuttle charge between the batteries [13], [14], [15]. This method of manipulation allows for the equalization of states such as state of charge (SOC) and temperature [16], [17], [18], [19], [20], as well as the output voltage [21], [22], [23], [24]. Another method of control is through the cooling of cells in a pack, or thermal control. Pack-level and cell-level arrangements can be used to balance cell temperatures, thereby improving performance and extending end-of-life (EOL) [25], [26].

When considering cell state control, an important step is analyzing the ability of the states to be observed and controlled. System analysis is well explored in the literature for observability due to its necessity for reliable state estimation [27], [28], [29], [30]. Several studies explore the difference between a linear and nonlinear observability analysis, finding that information can be lost when using a linearized model [31], [32]. As with any system property, however, the nonlinear metric has shown to be reliant on even small changes in battery parameters [33], [34], [35]. This indicates that the slight differences in new batteries from manufacturing tolerances are in fact capable of changing the results of analysis depending on the chosen cell for study. The conditioning of a battery's observability has been studied as well to determine the quality of estimates [36], and it has been shown that this can vary drastically between the linear approximation and true nonlinear system [37].

As for the controllability of lithium-ion battery states, the topic is investigated much less in literature, and many models are assumed controllable. The majority of the work related to control of lithium-ion batteries focuses on developing strategies for effective and efficient equalization [3], [4], [38], [39], [40]. Another sect of studies develops models which are more conducive to implementation in controllers [41], [42]. This is useful as it ensures engineers are able to identify and extract relevant information for more complex control strategies. With regard to the conduciveness of cells for control, however, limited studies perform a controllability analysis on the linearized version of the model [43]. This is done to ensure that a realized model retains the system properties other models are assumed to have.

There is currently a lack of insights into the controllability of battery models compared to those of observability. Current works that do study controllability rely on linear models or approximate models, which reduces the fidelity of the results. To strengthen the literature in these areas, this study provides two main contributions: (a) a correlation of the amount of control effort a cell requires as related to the condition number of its unique controllability matrix, and (b) a controllability and condition number analysis, performed for a set of experimentally parameterized batteries. The results of the analysis are then used to determine the best conditioned pack assemblies for a given batch of new and second-life cells.

To begin, this paper first describes the electrical dynamics of a battery through an equivalent circuit model in Section 2. Then, the calculation of nonlinear controllability and condition number are detailed and an example of the relationship between condition number and control effort is presented in Section 3. Next, the effects of parameter changes on mathematical cell conditioning at beginning- and end-of-life are explored in Section 4. Following this, a design scenario is introduced, and the condition numbers of new and aged cells are used to guide battery pack assembly in Section 5. Finally, this work is summarized and key insights are reviewed in Section 6.

## 2. BATTERY MODEL DESCRIPTION

This section describes an equivalent circuit model (ECM) for a lithium-ion battery. An ECM is a common way to represent a battery in a computationally efficient manner, consisting of several circuit components which represent the internal dynamics of the cell as seen in Figure 1. The bulk of the cell charge is captured in the nonlinear capacitor and is represented by the SOC, calculated as follows:

$$S\dot{O}C_j(t) = \frac{I_j(t)}{Q_j} \quad (1)$$

where $I_j$ is the input current going into cell $j$, and $Q_j$ is the capacity of the cell in coulombs.

The voltage of a lithium-ion battery relaxes from its value at the end of charge and discharge, which is captured by the series of resistor-capacitor (RC) pairs in Figure 1. The dynamics of the capacitors in these pairs are represented generally by (2), where $q_{i,j}$ and $\tau_{i,j}$ are the relaxation charge and time constant for RC pair $i$ respectively, and $N_{RC}$ is the number of RC pairs. These states combine with SOC into (3) to determine the cell voltage $V_j$, with $OCV_j$ representing the open-circuit voltage of the cell, $C_{i,j}$ as the relaxation capacitance for RC pair $i$, and $R_j$ as the cell's ohmic resistance. The parameters of the ECM that depend on SOC are considered to be polynomials whose coefficients vary between cells. These coefficients, as well as the cell capacities and number of RC pairs, are taken from the experimental parameterization results of two batches of lithium-iron phosphate batteries in [44], totaling to 66 batteries. The two batches are from different cell manufacturers and thus have slightly differing parameters, in spite of having the same chemistry. This will offer insight into the variance in control effort required of lithium-ion batteries.

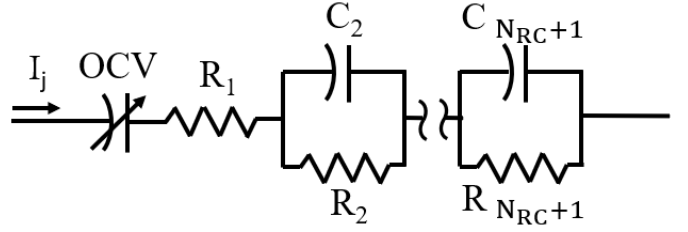

**FIGURE 1:** AN ECM REPRESENTING THE ELECTRICAL DYNAMICS OF A LITHIUM-ION BATTERY.

$$\dot{q}_{i,j}(t) = -\frac{1}{\tau_{i,j}(SOC_j(t))} q_{i,j}(t) + I_j(t) \quad (2)$$

$$V_j(t) = OCV_j(SOC_j) + \sum_{i=2}^{N_{RC}+1} \frac{1}{c_{i,j}(SOC_j(t))} q_{i,j}(t) + R_j(SOC_j(t)) I_j(t) \quad (3)$$

## 3. CONTROLLABILITY ANALYSIS

This section describes the conditions necessary for a nonlinear system to be controllable, as well as how to calculate the controllability and condition number. The condition number of the controllability matrix is then related to the amount of control effort required to drive the states of the nonlinear battery model to desired values.

### 3.1 Nonlinear Controllability

To determine controllability for a battery, the model described in the previous section is summarized in (4). Here, $f$ describes the nonlinear state dynamics which depend on the states, $h$ is the control-affine input relationship, and $x$ is the state vector, all of which are size $N_x \times 1$. Additionally, $u$ is the input vector of size $N_u \times 1$ and $\theta$ is the vector of capacity and parameter map polynomial coefficients of size $N_\theta \times 1$. A system is labeled controllable if its states can be driven from an arbitrary initial set of states, $x_0$, to an arbitrary final set of states, $x_f$, with a finite sequence of inputs $u \in [u_0, \dots, u_f]$.

$$\dot{x}_j = f(x_j, \theta_j) + h(\theta_j) u_j \quad (4)$$

The controllability of the continuous, nonlinear system of (4) is determined by assembling Lie brackets of the system into a matrix and determining its rank. The Lie algebra necessary to build this matrix is defined as follows:

$$ad_f^1 h = [f, h] = \frac{\partial h}{\partial x} f - \frac{\partial f}{\partial x} h$$

$$ad_f^2 h = [f,[f,h]] = \frac{\partial ad_f^1 h}{\partial x} f - \frac{\partial f}{\partial x} ad_f^1 h$$
$$\vdots$$
$$ad_f^k h = [f, ad_f^{k-1} h]$$

where $ad_f^k h$ is called the adjoint operator, the superscript of which indicates the bracket order. To determine controllability, the highest order $k$ is equal to one less than the dimension of $f$ and $h$, which is $N_x = 4$ for the experimental battery model presented in Section 2. Thus, the controllability matrix $C_o$ can be assembled from the Lie algebra of (4) into (5). For a given series of inputs, the system is said to be controllable about states $x_0$ if (5) is full rank when evaluated at $x_0$ [45]. This process can be repeated for any feasible state values to determine the broader scope of controllability for the system.

$$C_o = [h, ad_f^1 h, ad_f^2 h, ad_f^3 h] \quad (5)$$

By evaluating the appropriate Lie derivatives, the controllability matrix for (4) comes out to:

$$C_o =$$
$$\begin{bmatrix} \frac{1}{Q} & 0 & 0 & 0 \\ 1 & \frac{1}{\tau_2(x_1)} - \frac{\frac{\partial \tau_2}{\partial x_1} x_2}{Q\tau_2(x_1)^2} & \frac{1}{\tau_2(x_1)^2} - \frac{2\frac{\partial \tau_2}{\partial x_1} x_2}{Q\tau_2(x_1)^3} & \frac{1}{\tau_2(x_1)^3} - \frac{2\frac{\partial \tau_2}{\partial x_1} x_2}{Q\tau_2(x_1)^4} \\ 1 & \frac{1}{\tau_3(x_1)} - \frac{\frac{\partial \tau_3}{\partial x_1} x_3}{Q\tau_3(x_1)^2} & \frac{1}{\tau_3(x_1)^2} - \frac{2\frac{\partial \tau_3}{\partial x_1} x_3}{Q\tau_3(x_1)^3} & \frac{1}{\tau_3(x_1)^3} - \frac{2\frac{\partial \tau_3}{\partial x_1} x_3}{Q\tau_3(x_1)^4} \\ 1 & \frac{1}{\tau_4(x_1)} - \frac{\frac{\partial \tau_4}{\partial x_1} x_4}{Q\tau_4(x_1)^2} & \frac{1}{\tau_4(x_1)^2} - \frac{2\frac{\partial \tau_4}{\partial x_1} x_4}{Q\tau_4(x_1)^3} & \frac{1}{\tau_4(x_1)^3} - \frac{2\frac{\partial \tau_4}{\partial x_1} x_4}{Q\tau_4(x_1)^4} \end{bmatrix} \quad (6)$$

where the cell index $j$ is implied. Note that $C_o$ is full rank as long as the time constant maps are unique, which is true for the identified battery ECMs. This means that the battery model is controllable.

While satisfying the rank condition of controllability offers a binary answer to whether or not a system is controllable, there is still the issue of control feasibility. It is possible for a system to be theoretically controllable, yet, to achieve the desired final states, a large control effort is required due to the system being poorly mathematically conditioned [46]. This is determined by the matrix condition number of $C_o$, calculated according to:

$$\kappa(C_o) = \|C_o\| \|C_o^{-1}\| \quad (7)$$

where $\kappa(\cdot)$ is the condition number and $\|\cdot\|$ is the two-norm of the matrix. A matrix is considered to be well-conditioned for low magnitudes of $\kappa$ [47].

### 3.2 Control Effort Scenario

When managing a battery pack to meet performance demands, it is desirable to reduce the amount of control effort required. To describe the impact of electrical cell condition on this, consider two cells, A and B, with reduced dynamics from the battery model described in the previous section. Each battery has a bulk SOC and a single relaxation state. While both have an identical capacity of 4320 C, cell A has a relaxation state with a time constant of 10 seconds and cell B has one of 200 seconds. The voltage of the cells is calculated using a simplified, linear SOC-OCV map of $OCV_j(SOC_j) = 0.5 \times SOC_j + 3$ and a relaxation capacitance of 5000 C for both cells.

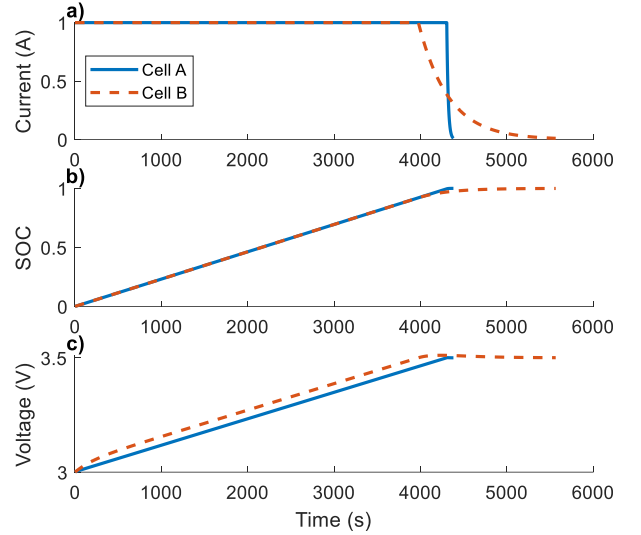

**FIGURE 2:** A COMPARISON OF (A) CURRENT, (B) SOC, AND (C) VOLTAGE DURING A CCCV CHARGE PROFILE FOR A BATTERY WITH A TIME CONSTANT OF 10 SECONDS (CELL A) AND ONE WITH A TIME CONSTANT OF 200 SECONDS (CELL B).

To determine the implication of this, a simulation of the two cells is run using the forward Euler method. The test consists of a constant current-constant voltage (CCCV) profile to charge the cells, a common test for updating parameter models while simultaneously meeting charge and discharge demands [48]. The cell SOCs start at 0%, and the relaxation states start at 0 C. As in a pack, the cells start the test receiving an equal current with a magnitude of $I_A = I_B = 1\ A$. Once a cell reaches the voltage limit of 3.5 V, its respective current relaxes until it drops below 10 mA. The results of this simulation can be seen in Figure 2.

Simulation shows that cell B takes 5,566 seconds to complete the profile, while cell A takes 4,382 seconds. This means cell B takes 1,184 seconds longer than cell A, even though it reaches the CV portion of the test sooner. This demonstrates, in an exaggerated manner, the effect of larger time constants on standard battery operation. If one cell has larger time constants than another cell, it will require more time to complete charging or discharging, which must be considered during pack management. If this extra time requires mediation, then a higher current, and thus higher maximum operating power, is necessary. The same principle is true for capacities, though larger charge capacities are typically regarded as beneficial due to a higher available energy.

It is understood that larger time constants will result in longer relaxation times, but it can be difficult to summarily quantify this increased effort as cell models grow more complex. As a proxy for control effort, the controllability condition number can succinctly represent the effect of these differences in parameters. Using the definition given by (7), the condition number can be calculated for cells A and B. This is done using the controllability matrix for a two-state battery model, which is given by (8). Doing so gives the condition numbers of $4.36 \times 10^4$ and $8.64 \times 10^5$ for cells A and B, respectively. The difference in the two values provides clarity to what was seen in simulation: *the battery with a higher condition number requires more control effort*.

$$C_o = \begin{bmatrix} \frac{1}{Q} & 0 \\ 1 & \frac{1}{\tau_2} \end{bmatrix} \quad (8)$$

## 4. ANALYSIS OF CONTROLLABILITY MATRIX CONDITION

This section presents an analysis of the controllability matrix condition number for the battery model described in Section 2. The conditioning of the cells is determined at the beginning-of-life (BOL), and the sensitivity of these values to changes in parameters is discussed. The effects of EOL parameters on conditioning are also explored, and the difference in controllability between BOL and EOL is described.

### 4.1 Beginning-of-Life

Calculating the condition number of a matrix analytically can prove difficult due to the challenge of determining the two-norm of large matrices. Therefore, the conditions for each experimentally parameterized cell model are calculated numerically for the four-state ECM. The condition number can be calculated about any appropriate trajectory for the battery, with these studies assuming equilibrium conditions for simplicity. Doing so gives the results of Figure 3, where the log of the mean cell's condition number as it varies with SOC is presented. Alongside this is the range of one standard deviation outside of the mean, represented by the shaded regions. The results are distinguished by cell manufacturers, with each group labeled batch 1 and batch 2 to isolate potential insights unique to one group.

The two batches have a clear similarity in mathematical conditioning, which is that each cluster of values dips around 60% SOC, then rises to a peak at roughly 77% before dropping off as SOC continues to increase. The difference between the two groups, however, highlights the necessity of analyzing their properties. The results of this indicate that, especially at lower SOCs, batch 1 will require more control effort than batch 2, which is useful to know during testing and pack assembly. The values in Figure 3 are not static. Battery properties vary from cell to cell as well as over time, as is the topic of much of this work. To gather some insight into how the mathematical conditioning of the cells will change with the parameters, a sensitivity analysis is performed. For conciseness, the mean cell of each batch will be the focus of this study. The sensitivity of $\kappa$ for each batch is determined by perturbing a parameter in (6) and dividing the change in value by the change in parameter. This value is then normalized by multiplying it by the original parameter as done in [49], [50]. The full calculation for this normalized sensitivity, $S_\theta$, is given by (9). To measure this, each parameter or parameter map that appears in (6) is independently increased by 1% of its nominal value.

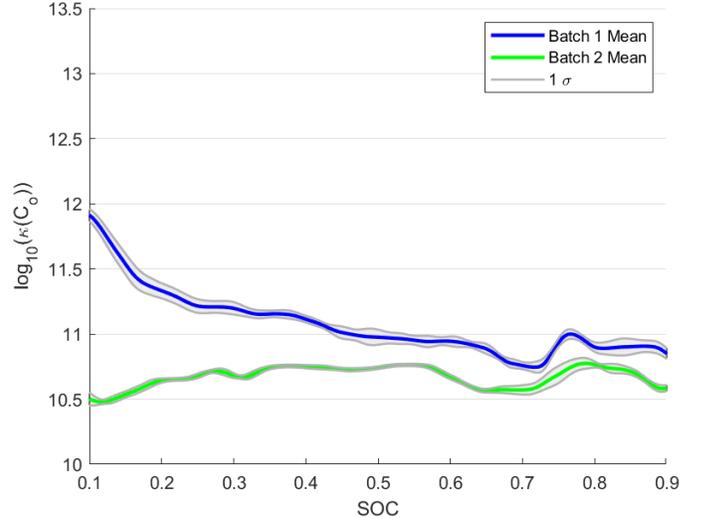

**FIGURE 3:** THE LOG OF THE MEAN CONDITION NUMBERS FOR EACH BATCH OF EXPERIMENTALLY PARAMETERIZED CELLS AS THEY CHANGE WITH RESPECT TO SOC.

$$S_\theta = \frac{\partial \kappa}{\partial \theta} \theta \quad (9)$$

The results of calculating sensitivity by disturbing the system parameters seen in Figures 4 and 5 reveal a major insight: *the condition number for the lithium-ion battery controllability metric is equally sensitive to all parameters*. Though initially surprising, it is expected that the relaxation parameters affect the condition number equally on a normalized scale. Due to the relaxation states' identical dynamics, aside from the magnitude of the time constant, it can be reasoned that any system properties related to them would be similarly affected by their parameters. Figures 4 and 5 show that there is a similar trend for capacity. This relationship is less intuitive, though there is a clear likeness between the sensitivities for each parameter type.

In lieu of an analytical solution, an example is used to understand the similar effects of the capacity and time constants on mathematical cell condition. The eigenvalues of (8) from the two-state model and its inverse when multiplied by their respective transpose are presented in (10). This is helpful for analyzing the condition number, which is calculated using the square roots of these terms. It can be seen in (10) that the capacity and time constant have matching terms in both expressions. This offers some clarity, as the eigenvalues show that a change to either parameter of equal amount will have a similar effect on the values. It is reasonable to assume this will extend to the higher order system, which explains the relationship between the sensitivity to capacity and time constants seen in Figures 4 and 5.

$$\lambda_{c_o,1,2} = \frac{\pm\sqrt{(Q^2\tau_2^2+Q^2-2Q\tau_2+\tau_2^2)(Q^2\tau_2^2+Q^2+2Q\tau_2+\tau_2^2)}}{2Q^2\tau_2^2}$$
$$+ \frac{1}{2Q^2} + \frac{1}{2\tau_2^2} + \frac{1}{2} \quad (10a)$$

$$\lambda_{c_o^{-1},1,2} = \frac{\pm\sqrt{(Q^2\tau_2^2+Q^2-2Q\tau_2+\tau_2^2)(Q^2\tau_2^2+Q^2+2Q\tau_2+\tau_2^2)}}{2}$$
$$+ \frac{Q^2}{2} + \frac{\tau_2^2}{2} + \frac{Q^2\tau_2^2}{2} \quad (10b)$$

**4.2 End-of-Life**

The majority of changes in parameters that lithium-ion batteries will experience over their lifetime is due to degradation. Cell ECM parameters degrade over time; most notably are the capacity and ohmic resistance of the cell. It has also been shown experimentally through electrochemical impedance spectroscopy that the lithium-ion diffusion resistance $R_i$ increases dramatically over the life of a lithium-iron phosphate battery cycled at 1 C charge. This impacts the time constants $\tau_i$ for each RC pair, resulting in a value at the EOL three times greater than at the BOL. By scaling the experimentally obtained parameters to the expected EOL values according to literature ($Q_{EOL} = 0.8 \times Q_{BOL}, \tau_{i,EOL} = 3 \times \tau_{i,BOL}$) [48], [51], [52], insights can be gathered about how cell conditioning changes over time.

The influence of aging on the control effort required for the cells is immediately clear when analyzing the new condition number distributions. Each batch's $\kappa$ increased by $3.02 \times 10^{12}$ and $9.81 \times 10^{11}$ relative to the BOL values, respectively. This makes sense, as aged cells are already expected to hinder management efforts by way of hitting limits early and shifting safe operating ranges. Comparing Figures 3 and 6 adds to this understanding, showing that aged batteries also inherently require more effort to control than fresh ones. The contribution of the parameters to this increase can be inferred from the BOL sensitivity analysis. The equal sensitivities indicate that the decrease in condition number due to the 20% reduction of capacity is overpowered by the increase caused by the 300% growth of each time constant. This means that the degradation of the lithium-ion diffusion resistance is the dominant mechanism causing the cells to require more control effort.

To determine how the condition number sensitivity changes as the cells age, the study from the previous subsection is repeated for the cells with aged parameters. There are two insights gained from this analysis:
- The controllability of a battery is equally sensitive to its capacity and time constants, regardless of how aged its parameters are.
- As a cell ages, the controllability becomes more responsive to changes in parameters when compared to the initial sensitivity.

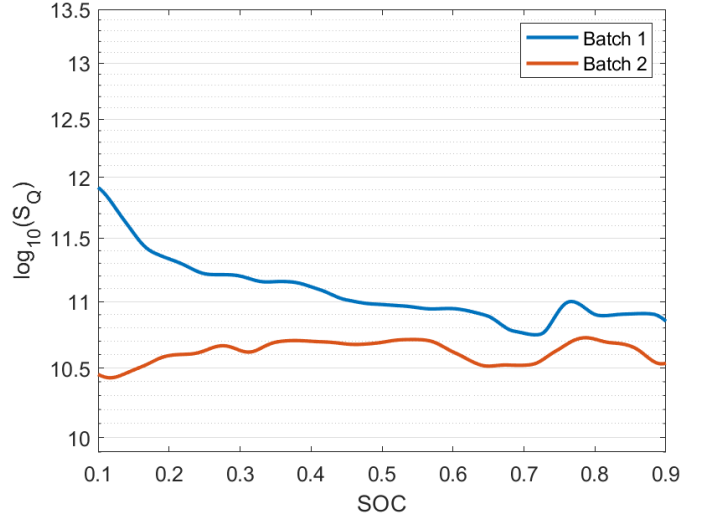

**FIGURE 4:** THE SENSITIVITY OF THE MEAN CONDITION NUMBER TO CAPACITY AS IT VARIES WITH SOC.

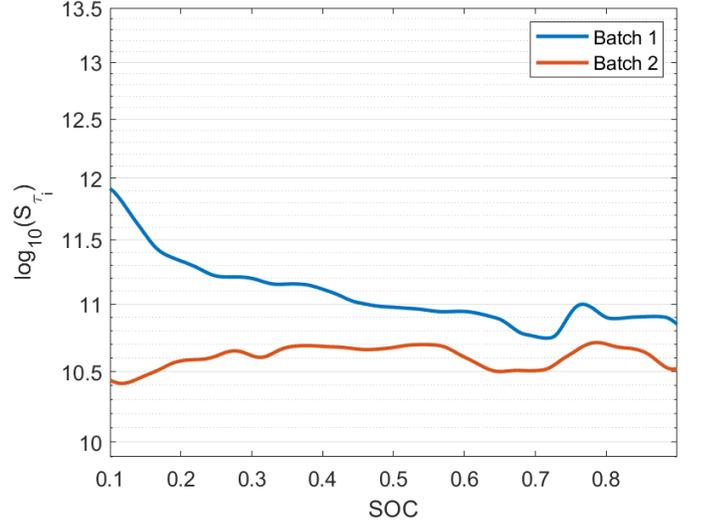

**FIGURE 5:** THE SENSITIVITY OF THE MEAN CONDITION NUMBER TO EACH TIME CONSTANT AS IT VARIES WITH SOC.

The first result confirms the relationship explored in the previous subsection, which indicated that a change in either the capacity or a time constant should result in an equivalent change to the condition number. Figures 7 and 8 show that this is true at EOL like it is at BOL, which makes sense because the insights from (10) remain the same. The latter insight is apparent when comparing Figures 4 and 5 to Figures 7 and 8. The sensitivity of batch 1 to both the capacity and time constants increased by $3.02 \times 10^{12}$, and batch 2's sensitivities increased by $8.55 \times 10^{11}$. This means that in addition to an aged battery requiring more control effort, variations in parameters will cause the amount required for each cell in a population to shift more drastically. The effect of this would worsen if heterogeneity in a pack is left unaddressed. The identical changes in sensitivity to each parameter also support the first point, emphasizing that all parameters play an equal role in determining the controllability of a cell.

## 5. CONTROLLABILITY ANALYSIS-AIDED DESIGN

Now that a general understanding of how the controllability of lithium-ion batteries changes over their life has been obtained, it can be used to guide the design of battery pack assemblies. This becomes more relevant as second-life cells strive to be used in commercial applications [53], [54]. It is clear from the previous analysis that aging has a drastic effect on the controllability of a cell, and thus the effort required to control its states. That introduces a new goal when selecting cells to be implemented into a pack: to maximize the effectiveness of control efforts made by a battery management system, the battery pack should be designed to have a low controllability condition number.

To study this, 32 of the tested cells are randomly selected to be aged, with slight variation, according to the EOL parameters used in the previous section. The variations for each chosen cell are intended to mimic those of a set of second-life batteries and can be seen in Figure 9. The remaining cells' parameters are fixed at their BOL values. From this population of healthy and aged cells, the objective is to divide the 66 cells into two battery pack assemblies. The MATLAB command `randperm` is used to obtain 10,000 design options for two packs of 33 cells randomly chosen from the population.

When selecting a set of assemblies from these options, there are two objectives of interest. The first is according to a standard design choice, which is to maximize the capacity of each pack to meet power demands. This is assessed by determining the smallest average pack capacity of each design and comparing them to each other. The best design is selected by finding the largest of these values. While capacity is an important metric to consider, it does not guarantee that the packs will be well conditioned for control. To determine how they compare, the second objective is to minimize the controllability condition number of each pack. By comparing the largest condition number of each design choice, the ideal choice based on this metric is the one with the smallest value. The results for the first 10 designs are displayed in Table 1, with the best design according to the respective metric marked with an asterisk.

The first 10 combinations demonstrate that the two objectives are not in sync with one another. There are two instances where each metric is at its maximum or minimum respectively, and none of them are for the same designs. This indicates that there is no obvious choice for the pack with the best capacity and controllability, and there does not appear to be a relationship between the two when analyzing the other designs.

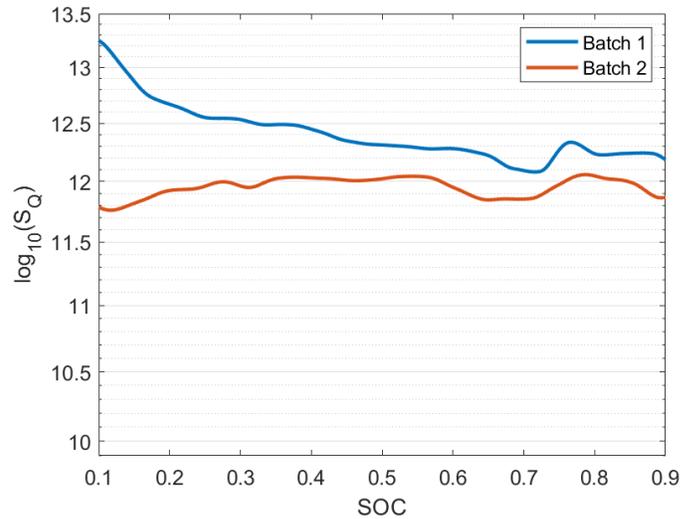

**FIGURE 7:** THE LOG OF THE SENSITIVITY OF THE MEAN CONDITION NUMBER TO CAPACITY AS IT VARIES WITH SOC, WITH PARAMETERS SCALED TO REFLECT EOL.

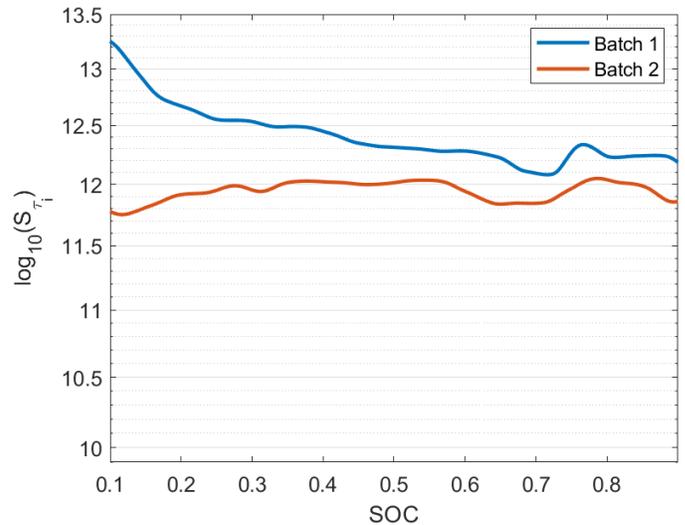

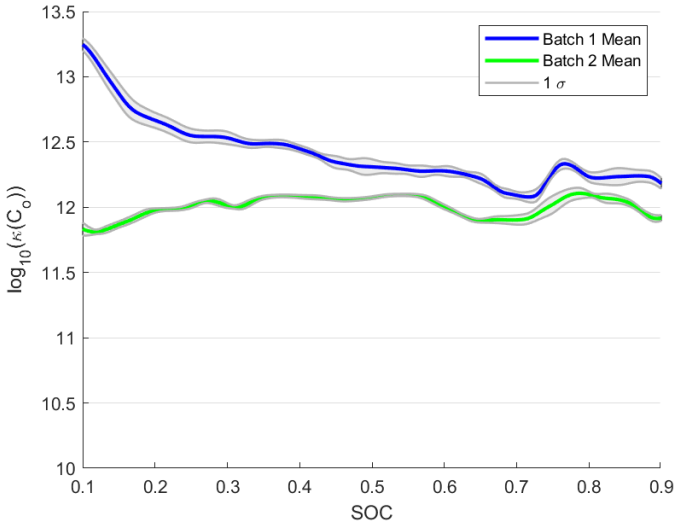

**FIGURE 6:** THE LOG OF THE MEAN CONDITION NUMBERS FOR EACH BATCH OF CELLS AS THEY CHANGE WITH RESPECT TO SOC, WITH PARAMETERS SCALED TO REFLECT EOL.

**FIGURE 8:** THE LOG OF THE SENSITIVITIES OF THE MEAN CONDITION NUMBER TO EACH TIME CONSTANT AS IT VARIES WITH SOC, WITH PARAMETERS SCALED TO REFLECT EOL.

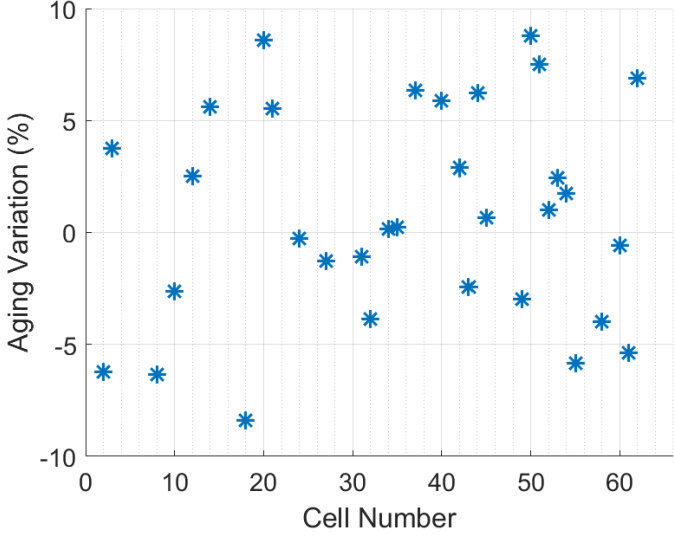

**FIGURE 9:** THE RANDOMIZED VARIATION IN AGING FOR SECOND-LIFE BATTERY PARAMETERS USED IN THIS WORK.

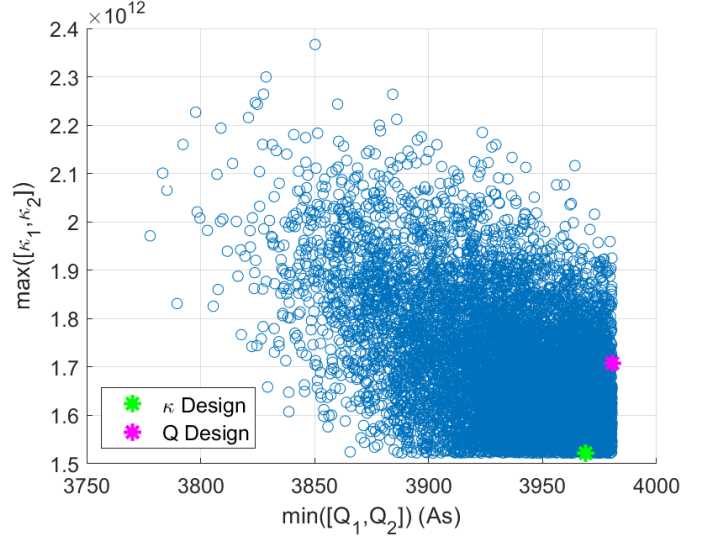

**FIGURE 10:** THE MAXIMUM CONDITION NUMBER OF EACH PACK AND THE MINIMUM CAPACITY OF EACH PACK FOR 10,000 DESIGN OPTIONS.

To explore this further, the two metrics described previously are plotted against each other for all 10,000 designs in Figure 10. The best options with respect to each metric are highlighted by a green and magenta marker to provide insight like in Table 1.

The results of the analysis for more combinations highlight one major thing: designing for a standard metric such as pack capacity is not always the same as designing for controllability. Figure 10 clearly shows that the best designs with regard to each of these is not in agreement. It also shows that there are numerous solutions which provide a high pack capacity but are poorly conditioned, and vice versa. This means that when designs are selected only with regard for capacity, there is a strong chance that the packs will require more control effort. There is also a chance that a chosen combination produces a value closer to the bottom right of Figure 10. However, to ensure the controllability of the battery packs are not left to chance, the conditioning of the cells' controllability matrix must be considered during design.

## 6. CONCLUSION

This work studies the controllability of lithium-ion batteries and how the control effort required for a cell changes with its parameters. The condition number of a cell's controllability matrix is linked to the amount of time necessary for management. A sensitivity study is performed for cells with both fresh and aged parameters, showing that cell controllability is equally affected by capacity and time constants. This idea is translated to design, and the distribution of ideal capacity and controllability of a pack is explored, showing that a more controllable pack is something which must be designed for.

## ACKNOWLEDGEMENTS

This material is based upon work supported by the National Science Foundation under Award No. 2324707.

**TABLE 1:** THE CAPACITY IN AH AND CONDITION NUMBER, ALONG WITH THEIR RESPECTIVE DESIGN METRICS AND HIGHLIGHTED IDEAL DESIGNS, FOR THE FIRST 10 RANDOMIZED COMBINATIONS.

| Design | $Q_{1,avg}$ | $Q_{2,avg}$ | $\min(Q_{i,avg})$ | $\kappa_{1,avg}$ | $\kappa_{2,avg}$ | $\max(\kappa_{i,avg})$ |
|---|---|---|---|---|---|---|
| 1 | 1.116 | 1.096 | 1.096 | $1.72e12$ | $1.33e12$ | $1.72e12$ |
| 2 | 1.111 | 1.100 | 1.100 | $1.43e12$ | $1.62e12$ | $1.62e12$ |
| 3 | 1.090 | 1.122 | 1.090 | $1.53e12$ | $1.51e12$ | $1.53e12^*$ |
| 4 | 1.065 | 1.147 | 1.065 | $2.07e12$ | $0.98e12$ | $2.07e12$ |
| 5 | 1.096 | 1.115 | 1.096 | $1.61e12$ | $1.43e12$ | $1.61e12$ |
| 6 | 1.102 | 1.110 | $1.102^*$ | $1.34e12$ | $1.70e12$ | $1.70e12$ |
| 7 | 1.115 | 1.096 | 1.096 | $1.51e12$ | $1.53e12$ | $1.53e12^*$ |
| 8 | 1.102 | 1.109 | $1.102^*$ | $1.62e12$ | $1.42e12$ | $1.62e12$ |
| 9 | 1.118 | 1.094 | 1.094 | $1.20e12$ | $1.84e12$ | $1.84e12$ |
| 10 | 1.116 | 1.095 | 1.095 | $1.64e12$ | $1.40e12$ | $1.64e12$ |

*The best design choice according to the design goal of the respective metric.